\documentclass[aps,prb,reprint,superscriptaddress]{revtex4-1}
\usepackage{graphicx}
\usepackage{fullpage}
\usepackage{amsmath,amssymb}
\usepackage{color}
\usepackage{setspace}
\usepackage{hyperref}
\usepackage{array}

 \makeatother

\begin{document}

\title{Valley polarization assisted spin polarization in two dimensions}

\author{V. T. Renard\footnote{Correspondence to : vincent.renard@cea.fr}}
\affiliation{Univ. Grenoble Alpes/CEA, INAC-SPSMS, F-38000 Grenoble, France}

\author{B. A. Piot}
\affiliation{Laboratoire National des Champs Magn\'{e}tiques Intenses, CNRS-UJF-UPS-INSA, 38042 Grenoble, France}

\author{X. Waintal}
\affiliation{Univ. Grenoble Alpes/CEA, INAC-SPSMS, F-38000 Grenoble, France}

\author{G. Fleury}
\affiliation{Service de Physique de l'\'{E}tat Condens\'{e}, DSM/IRAMIS/SPEC, CNRS UMR 3680 CEA Saclay, 91191 Gif sur Yvette cedex, France }

\author{D. Cooper}
\affiliation{Univ. Grenoble Alpes/CEA Leti Minatec campus, F-38054 Grenoble, France}

\author{Y. Niida}
\affiliation{Graduate School of Science, Tohoku University, 6-3 Aramakiaza Aoba, Aobaku, Sendai, 980-8578 Japan}

\author{D. Tregurtha}
\affiliation{Department of Physics, University of Bath, Bath BA2 7AY, UK}

\author{A. Fujiwara}
\affiliation{NTT Basic Research Laboratories, NTT Corporation, Atsugi-shi, Kanagawa 243-0198, Japan}

\author{Y. Hirayama}
\affiliation{Graduate School of Science, Tohoku University, 6-3 Aramakiaza Aoba, Aobaku, Sendai, 980-8578 Japan}

\author{K. Takashina}
\affiliation{Department of Physics, University of Bath, Bath BA2 7AY, UK}

\date{\today}

\maketitle


Valleytronics is rapidly emerging as an exciting area of basic and applied research. In two dimensional systems, valley polarisation can dramatically modify physical properties through electron-electron interactions as demonstrated by such phenomena as the fractional quantum Hall effect and the metal-insulator transition. Here, we address the electrons' spin alignment in a magnetic field
in silicon-on-insulator quantum wells under valley polarisation. In stark contrast to expectations from a non-interacting model, we show experimentally that less magnetic field can be required to fully spin polarise a valley-polarised system than a valley-degenerate one. Furthermore, we show that these observations are quantitatively described by parameter free \textit{ab initio} quantum Monte Carlo simulations. We interpret the results as a manifestation of the greater stability of the spin and valley degenerate system against ferromagnetic instability and Wigner crystalisation which in turn suggests the existence of a new strongly correlated electron liquid at low electron densities.

\section*{Introduction}

The valley degree of freedom has a long history as a subject of pure and applied research since it is an intrinsic property of the band structure of silicon and germanium, the historical materials in microelectronics\cite{AndoFowlerStern1982}. Valley degeneracy had generally been viewed as a drawback as it limits the mobility of CMOS (Complementary Metal Oxide Semiconductor) devices due to intervalley scattering\cite{Schaffler1997}. Microelectronics manufacturers have consequently put much effort into  manipulating valley bands through strain, to improve transport properties. This approach has been successful, and strained silicon has been in use in microelectronics since the 90 nm node\cite{Ghani2003}.

More recently, however, the valley degree of freedom is becoming recognised as an opportunity, rather than a hindrance, and this is leading to the emergence of a field of research now known as valleytronics in which valleys are exploited in addition to charge and spin. Valleytronics has received a recent boost owing to the discovery of graphene and other new topical materials also possesing the valley degree of freedom and by the proposal of valleytronics devices\cite{Gunawan2006,Beenakker2007,Culcer2012,Behnia2012,Isberg2013,Jones2013,Xu2014}. A vital ingredient to the development of valleytronics is valley polarization. Analogous to spin polarization in spintronics in which when achieved under equilibrium conditions, leads to key phenomenology such as ferromagnetism, valley polarization can also be expected to yield rich and useful physics.

Experimental research into the physical consequences of valley polarizing a two dimensional electron system in the steady state is led by studies of AlAs and Si based structures. It has been demonstrated that valley polarisation dramatically affects phenomena such as the fractional quantum Hall effect\cite{Shkolnikov2005,Bishop2007,Padmanabhan2010,Gokmen2010} and the metal insulator transition\cite{Shayegan2007,Takashina2011,Takashina2013,Renard2013}, two effects in which electron-electron interactions are central. Pioneering experiments performed in AlAs indicate that valley polarisation also has a strong impact on another effect where electron-electron interactions play crucial roles : spin polarisation. It has been demonstrated that valley polarisation leads to a strong enhancement of spin susceptibility and symmetrically, spin polarisation enhances valley susceptibility \cite{Shayegan2004,Shayegan2008,Shayegan2010}.

In this article, we firstly confirm the enhancement of spin susceptibility by valley polarisation in silicon, which in contrast to AlAs has an isotropic in-plane effective mass which simplifies interpretation of transport phenomena. More importantly, we explore a new regime in the interaction-disorder parameter space where a qualitatively new behaviour emerges. This is achieved by using electrically controlled valley polarisation in a simple two dimensional electron gas in foundry compatible silicon-on-insulator (100) MOSFETs\cite{Takashina2004,Takashina2006} (Metal Oxide Semiconductor Field Effect Transistors). We present magneto-resistance data which indicate that for low enough electron densities, valley polarising the  two-dimensional electron gas (2DEG) reduces the field of full spin polarisation. This represents not only a quantitative failure of the single particle picture but a qualitative one, in which the observed behaviour is opposite to the prediction of the non-interacting framework. 

\begin{figure}[h]
\includegraphics[width=\columnwidth]{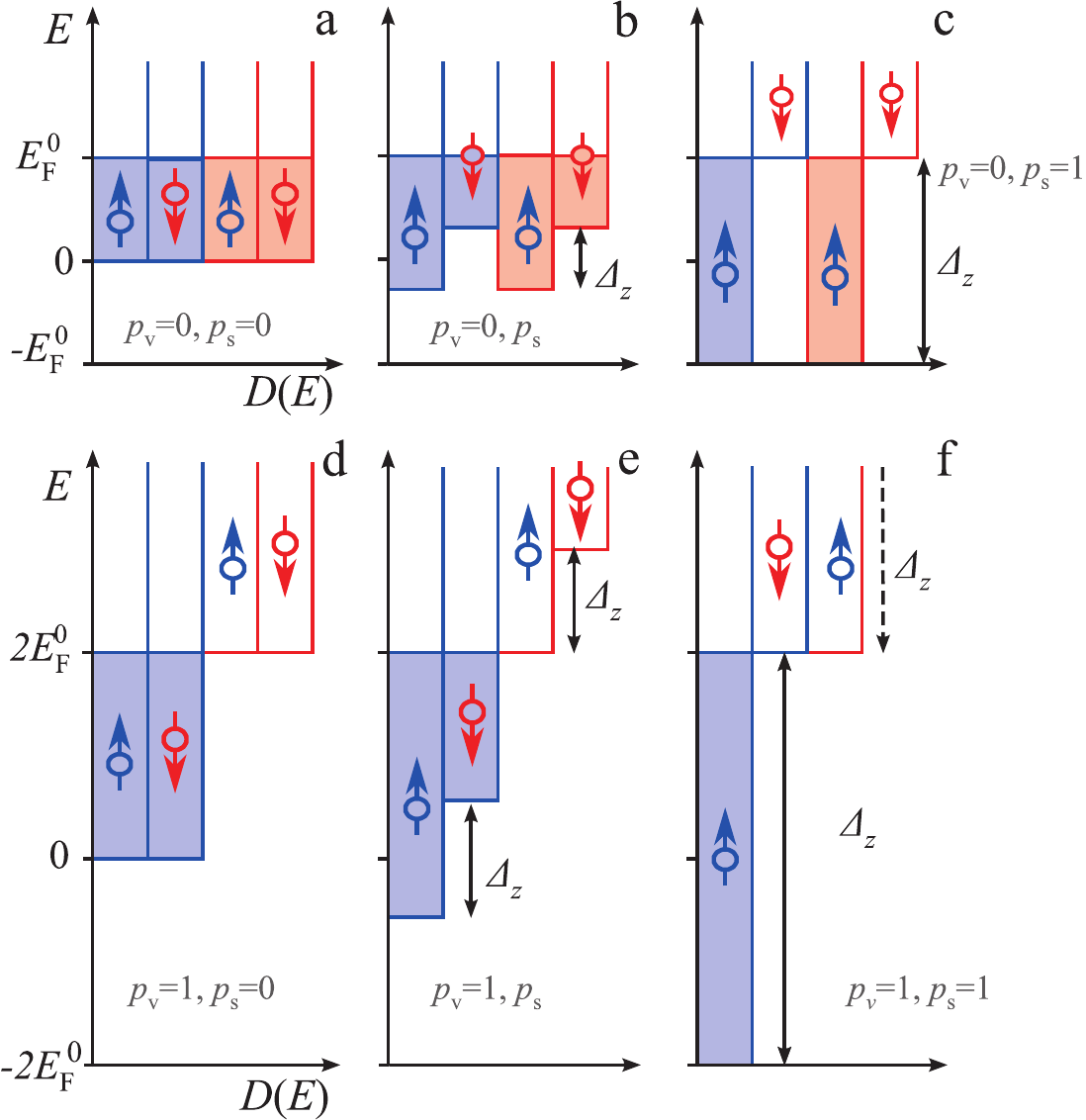}

\caption{\textbf{Energy diagram of a 2D electron gas of fixed electron density}. The diagram depicts the four spin-split valley-split subbands. Up (resp. down) arrows correspond to spin up (down) electrons and blue (resp. red) color corresponds to valley + (resp. -). Depending on spin and valley polarisation, the system can be either (a) valley- and spin-degenerate 
or (b) valley-degenerate, partially spin-polarised or (c) valley-degenerate, spin-polarised or (d) valley-polarised, spin-degenerate or (e) 
valley-polarised, partially spin-polarised or (f) valley-polarised, spin-polarised. \label{Simple_Band}}
\end{figure}

\begin{figure*}
\includegraphics[width=2\columnwidth]{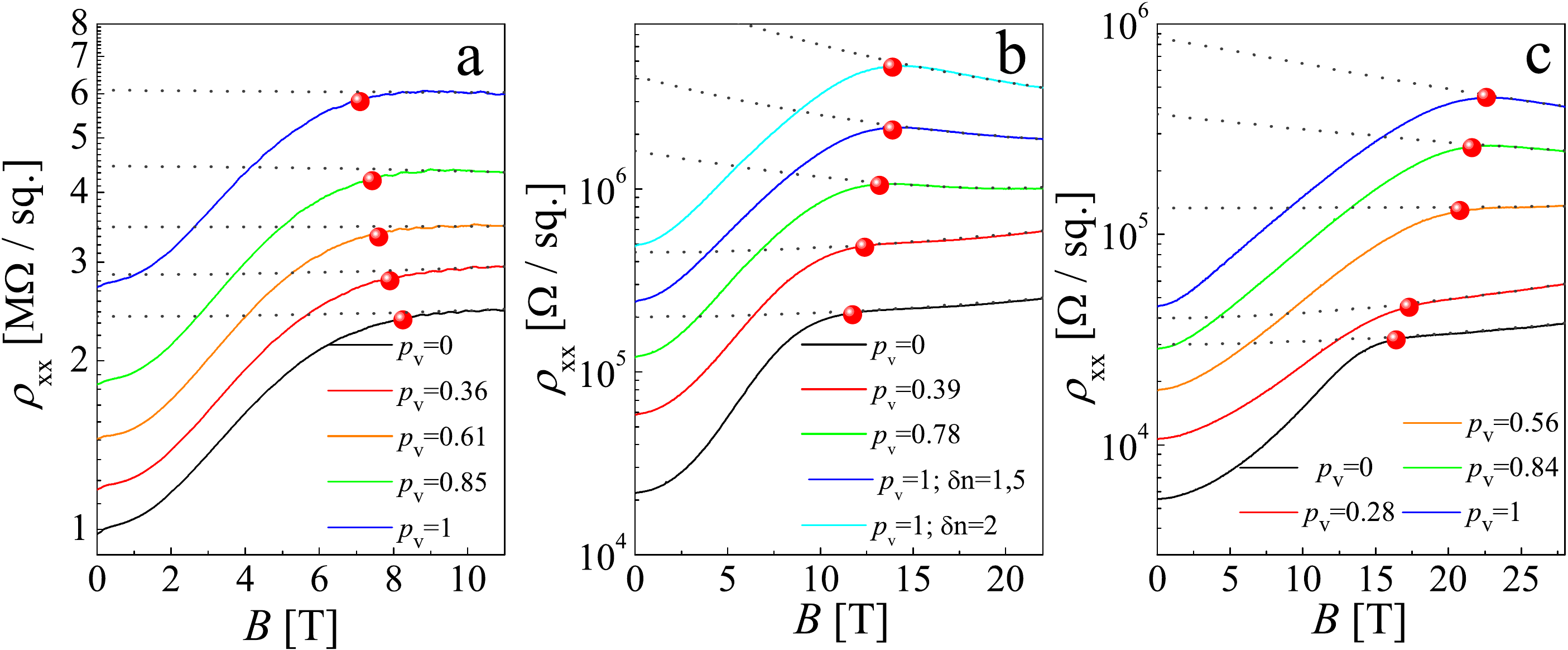}

\caption{\textbf{Magneto-resistance as function of valley polarisation.} The MR of the 2DEG is plotted for increasing $p_\textrm{v}$ at the density of $n=1.6\times10^{15}$ m$^{-2}$ (a), $n=2.5\times10^{15}$ m$^{-2}$ (b) and $n=3.5\times10^{15}$ m$^{-2}$ (c), at $T=1.65$ K (See Fig.~\ref{negative_delta} for a version of Fig.~\ref{Valley_dep}a in linear scale). The indicated values for $p_\textrm{v}$ correspond to those at full spin polarisation. The field of full spin polarisation (plotted as red bullets) is estimated from the field where the MR has reached 97.5$\%$ of its high-field spin-independent behaviour (grey dotted lines). Values of $\delta n$ are given in units of 10$^{16}$ m$^{-2}$ in panel b. \label{Valley_dep}}
\end{figure*}

\section*{Results}
\subsection*{Single particle picture}
Let us first describe briefly how spin polarization is expected to respond to valley polarisation in a non-interacting 2DEG.
In the non-interacting model, the density of states of a 2DEG is independent of energy and can be written as $g_\textrm{s}g_\textrm{v}D_\textrm{0}$ where $D_\textrm{0}=m_\textrm{b}/2\pi\hbar^2$, $g_\textrm{s}$ and $g_\textrm{v}$ are spin and valley degeneracies, $m_\textrm{b}$ is the electron band mass and $\hbar$ is the reduced Planck's constant. In a (001) silicon 2DEG, $g_\textrm{s}=g_\textrm{v}=2$ and the system is composed of four independent spin-valley subbands with equal density of states $D_\textrm{0}$, as depicted in Figure~\ref{Simple_Band}a. All states are filled up to the Fermi energy $E_\textrm{F}^\textrm{0}= n/g_\textrm{s}g_\textrm{v}D_\textrm{0}\equiv n/4D_\textrm{0}$ at zero temperature, where $n$ is the electron sheet density.

 Applying a magnetic field parallel to the electron gas raises the bottom of the spin down bands compared to that of the spin up bands by the Zeeman splitting $\Delta_z=g\mu_\textrm{B}B$ (Fig.~\ref{Simple_Band}b) where $g$ is the Land\'{e} g-factor ($g=2$ in Si) and $\mu_\textrm{B}$ the Bohr magneton. Spin down electrons are consequently transferred to the spin up band: the system spin polarises. The spin (valley) polarisation is defined as $p_\textrm{s}=(n_\uparrow-n_\downarrow)/n$ (resp. $p_\textrm{v}=(n_+-n_-)/n$) with  $n_\uparrow$ and $n_\downarrow$ being the spin up and spin down electron densities respectively ($n_+$ and $n_-$ are the electron densities in the $+$ and $-$ valleys). At $p_\textrm{v}=0$, full spin polarisation is achieved when $\Delta_z=2E_\textrm{F}^\textrm{0}$ (Fig.~\ref{Simple_Band} c). It follows that the field $B_\textrm{p}$ required for full spin polarisation should double when the system is valley polarised because the kinetic energy in the valley polarised system (Fig.~\ref{Simple_Band}d) is twice that of the unpolarised one (Fig.~\ref{Simple_Band}a). For more details on this single particle picture see Supplementary Discussion.
 
\subsection*{Experimental determination of $B_\textrm{p}$}
The field of full spin polarisation $B_\textrm{p}$ of a 2DEG can be extracted from measuring the electrical resistance under in-plane magnetic field. Spin polarisation has the effect of reducing the ability of the 2DEG to screen disorder, and as a consequence, increases the resistance due to enhanced scattering until, in the simplest case, the magneto-resistance (MR) saturates to a constant value\cite{Gold2000}. This is a signature that full spin polarisation is reached and the spin degree of freedom is completely frozen.

 However, this situation is only rarely observed in experiments where it is more common to observe a shoulder in the magnetoresistance\cite{Okamoto1999,Tutuc2001,Shashkin2001,Pudalov2002,Shayegan2004,Lai2005,Boukari2006,Piot2009,Kapustin2009} when the spin system freezes but the resistance continues to change due to spin-independent effects such as the coupling of the magnetic field to the electrons' orbital motion\cite{DasSarma2000,Piot2009}. \\
\indent In the absence of a comprehensive description of the high field behaviour, we follow previous literature in empirically estimating $B_\textrm{p}$ as the field where the MR has reached 97.5~$\%$ of its high field dependence\cite{Pudalov2002,Shayegan2004} (The behaviour in the spin polarized regime at high fields was fitted to a quadratic behaviour with no particular physical meaning. See dotted lines in Fig.~\ref{Valley_dep}). Results are shown as red dots in Fig.~\ref{Valley_dep}.\\
\indent We note that we have also estimated $B_\textrm{p}$ following other methods used in the literature~\cite{Pudalov2002} (inflexion point in the MR, intersection of the high field asymptote and tangent at the inflexion point etc.). All methods provided a qualitatively similar behaviour in the changes of $B_\textrm{p}$ with valley polarisation. We have chosen the method described above as we believe it provides a better quantitative estimation of $B_\textrm{p}$ compared to other methods which underestimate it.

\subsection*{Evolution of $B_\textrm{p}$ with valley polarisation}
At large density (Fig.~\ref{Valley_dep}b and c), we observe an increase of $B_\textrm{p}$ with valley polarisation, a behaviour which is qualitatively consistent with the single particle picture (See Refs.~\onlinecite{Takashina2004,Takashina2006} and Methods for details on the electrical control and determination of $p_\textrm{v}$). It should be stressed, however, that quantitatively, the single-particle model fails completely. The values of $B_\textrm{p}$ are always much lower than expected. Also, instead of the doubling of $B_\textrm{p}$ from $p_\textrm{v}=0$ to $p_\textrm{v}=1$, we observe only a moderate increase of $B_\textrm{p}$. For exemple, we measure $B_\textrm{p}(p_\textrm{v}=0)=11.75$ T and $B_\textrm{p}(p_\textrm{v}=1)=13.9$ T for $n=2.5\times10^{15}$ m$^{-2}$ while the single particule picture predicts $B_\textrm{p}(p_\textrm{v}=0)=28.85$ T and $B_\textrm{p}(p_\textrm{v}=1)=51.7$ T respectively. These observations are consistent with the strong enhancement of spin susceptibility with valley polarisation seen in AlAs\cite{Shayegan2004, Shayegan2008,Shayegan2010}. We note in addition, that the quantitative failure of the single-particle model has also been observed in numerous experiments where valley polarisation could not be tuned, regardless of valley degeneracy \cite{Okamoto1999,Tutuc2001,Shashkin2001,Pudalov2002,Lai2005,Boukari2006,Piot2009,Kapustin2009}. 

At lower density (Fig.~\ref{Valley_dep}a), the single particle picture fails qualitatively. Here, our data show that $B_\textrm{p}$ moves to lower and lower magnetic field as valley polarisation is increased. That is, it becomes easier to spin polarise a valley-polarised electron gas than a valley degenerate one at low enough density.

\begin{figure}[h]
\includegraphics[width=\columnwidth]{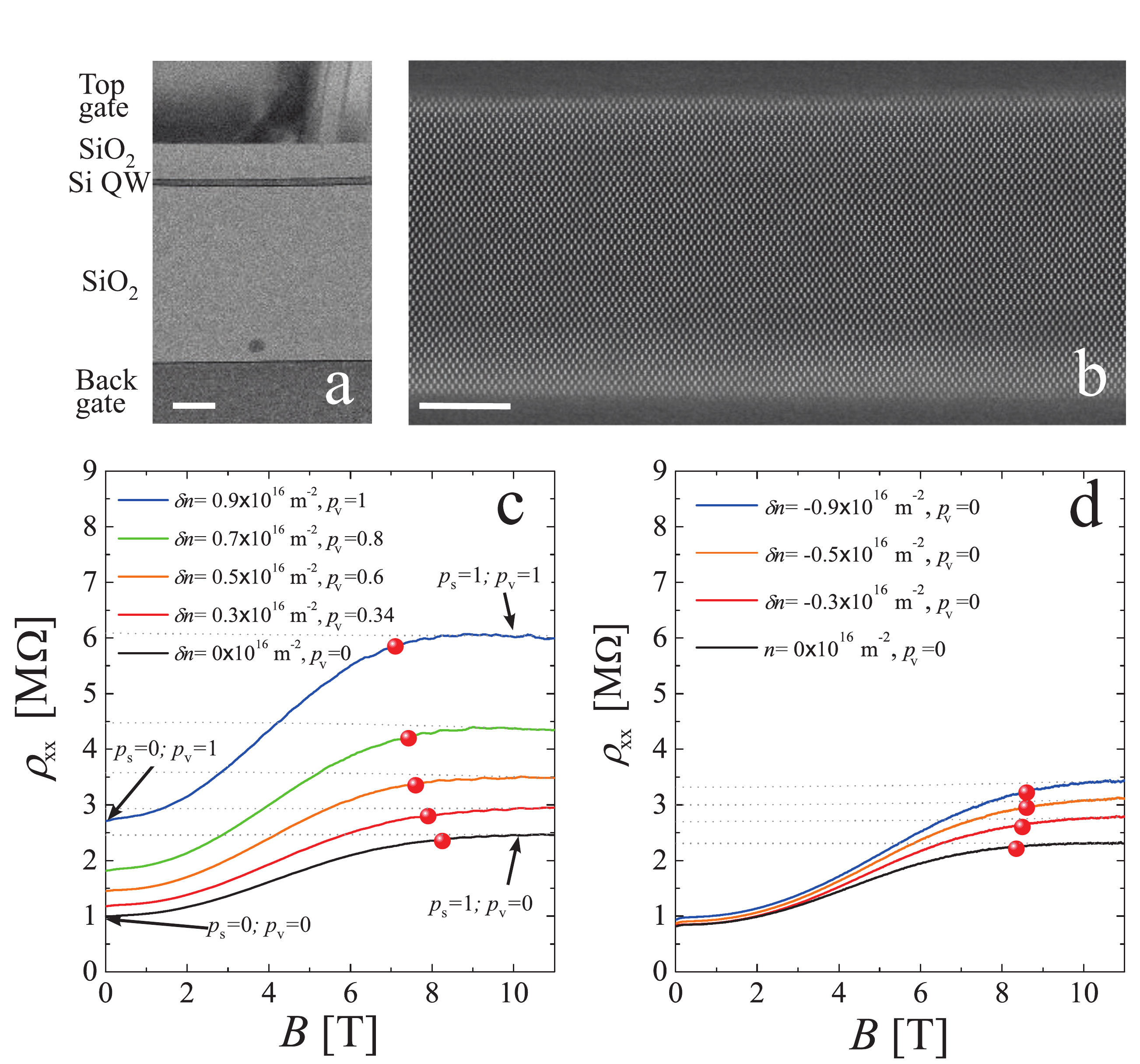}

 \caption{\textbf{Exclusion of disorder as the origin of the reduction in $B_\textrm{p}$.} a) Bright field scanning TEM cross-section of a 16 nm SIMOX device made with the same recipe as our sample. The scale bar is 100 nm b) High Angle Annular Dark Field scanning TEM image of the quantum well. The image shows the very good and similar crystallographic quality of the top and buried interfaces. The scale bar is 5 nm c) Electrons are pressed against the back interface $\delta n>0$. d) Electrons are pressed against the front interface for $\delta n<0$. Red dots mark the field of full spin polarisation in both panels.  Here, $n=1.6\times10^{15}$ m$^{-2}$ in both panels while $T=1.6$ K on the panel c and $T=1.7$ K in the panel d. (This explains the small difference in resistivity at $B=0$ T and $\delta n=0$.) The grey dashed lines correspond to the high-field behaviour used for the determination of $B_\textrm{p}$. \label{negative_delta}} 
\end{figure}

\subsection*{Exclusion of disorder as the cause of the reduction of $B_\textrm{p}$ with valley polarisation}
One cause for this observation could be an increase in disorder which is known to reduce $B_\textrm{p}$\cite{Pudalov2002}. At first sight, this explanation might seem plausible since valley polarisation is achieved by pressing the 2DEG against the buried Si/SiO$_2$ interface (see Methods) and hence the 2DEG experiences higher disorder due to interface roughness. The amplitude of this effect can be estimated in our device where transport can also be investigated at the top Si/SiO$_2$ interface. At this interface, valley splitting is negligible\cite{Takashina2006} but disorder is comparable as demonstrated by our high resolution Transmission Electron Microscope (TEM) images (Fig.~\ref{negative_delta}) and independent measurements with holes which do not possess the valley degree of freedom and show similar mobility at both interfaces\cite{Niida2013}. This enables us to separate the effects of disorder and valley polarisation which are mixed at the buried interface.

Figures~\ref{negative_delta}c and d show the magnetoresistance of the 2DEG for comparable magnitudes of out-of-plane electric bias (estimated from the phenomenological parameter $\delta n$. See Methods) when electrons are pressed against the buried or front Si/SiO$_2$ interface. The magnitude of the bare disorder potential increases with $\left|\delta n \right|$, while valley polarisation is enhanced only for $\delta n >0$.\cite{Takashina2006}
Insight into the variation of the bare disorder with $\delta n>0$ is found from the comparison of the resistance at ($p_\textrm{s}=1; p_\textrm{v}=0$) and ($p_\textrm{s}=0; p_\textrm{v}=1$) (highlighted in Fig.~\ref{negative_delta}c). 

Spin and valley degeneracy can be treated as formally equivalent in the ``screening'' description of transport described in Ref.~\onlinecite{Gold2000}. Therefore, in the absence of variation of disorder due to the process of polarising, one should expect the same increase of resistance due to equivalent reduction of screening regardless of which degeneracy is lifted\cite{Takashina2013}. 

\begin{figure}[t!]
\includegraphics[width=\columnwidth]{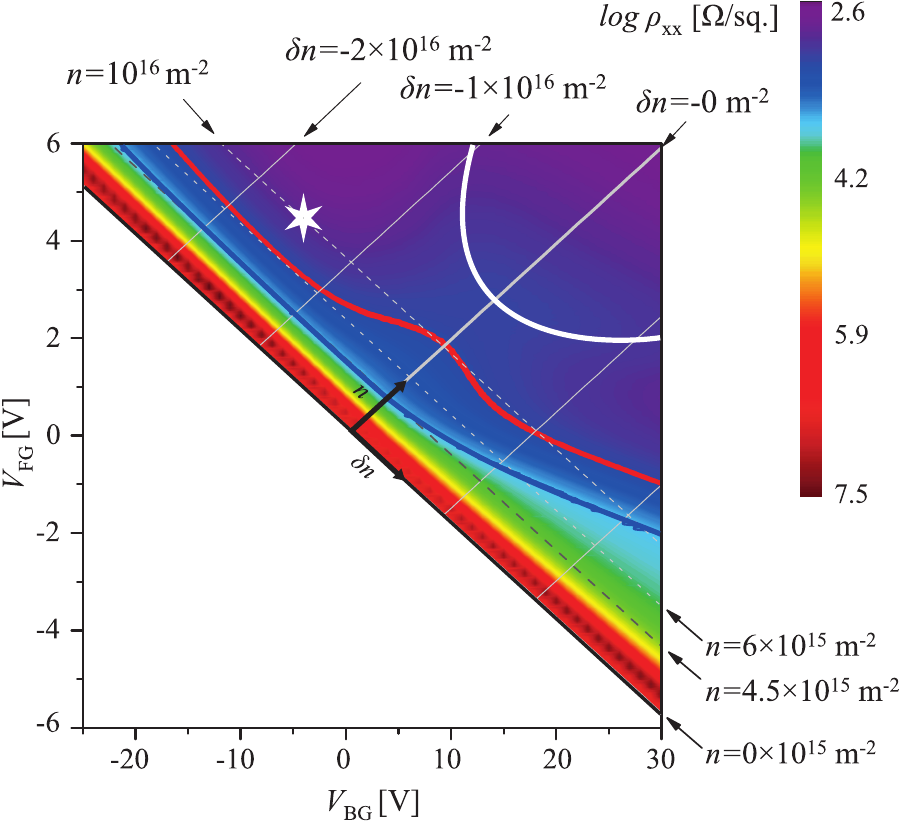}

\caption{\textbf{Two dimensional map of the resistance.} Resistance of the sample as function of front and back gate voltages in a log scale. The white line materialises the upper spatial subband edge.
The red curve represents the $\rho_{xx}=1600 \Omega$ iso-resistance obtained for $(n=10^{16} $m$^{-2},\delta n=0$ m$^{-2})$ while the blue one represents the $\rho_{xx}=6000 \Omega$ iso-resistance obtained for $(n=4.5\times10^{15} $m$^{-2},\delta n=0$ m$^{-2})$. The mobility was measured at the white star (see Methods for more details).  \label{Fig4}} 
\vspace{.25cm}
\end{figure}

We find that the resistances are indeed almost the same at ($p_\textrm{s}=1; p_\textrm{v}=0$) and ($p_\textrm{s}=0; p_\textrm{v}=1$). Therefore, we conclude that the major part of the increase in resistance with $\delta n>0$ seen in Fig.~\ref{negative_delta}a at $B$=0~T can be attributed to a reduced screening due to valley polarisation and not to a significant increase of bare disorder. Furthermore, the increase of disorder with $\delta n>0$ can be estimated to be about 10\% from the observed 10\% difference between the resistance at ($p_\textrm{s}=0; p_\textrm{v}=1$) and ($p_\textrm{s}=1; p_\textrm{v}=0$). This estimation of the variation of disorder with $\lvert\delta n\lvert$ is confirmed by transport at the front interface where the entire change in resistance must be attributed to a change in disorder. The data at $B=0$~T in Fig.~\ref{negative_delta}d reveal a 14 \% increase in the resistance for $\delta n=-0.9\times10^{16}$ m$^{-2}$ (comparable in amplitude to that necessary for full valley polarisation at the buried interface). The weak dependence of disorder on $\delta n<0$ is well illustrated in the resistance map shown in Fig.~\ref{Fig4} where contours of constant resistance run parallel to the constant density lines in the relevant regime of density. 

Importantly, Fig.~\ref{negative_delta}d shows that the field of full spin polarisation is almost unchanged when electrons are pressed against the front interface. That is, the 14\% change in the bare disorder with $\delta n$ in this experiments is not sufficient to cause a substantial change in $B_\textrm{p}$. We even note a small initial increase of $B_\textrm{p}$ which might be attributed to the removal of a small valley splitting at symmetry ($\delta n = 0$) as electrons are moved away from the buried interface. Alternatively, this small increase could be due to a broadening of the spin band edge due to the increased disorder felt by the 2DEG. Either way, we can conclude from this that the increase of the bare disorder with $\lvert\delta n\lvert$ cannot explain the behaviour seen in Fig.~\ref{Valley_dep}a, leaving valley polarisation as the only culprit. 

As a final confirmation, we have checked that increasing $\delta n$ above full valley polarisation does not lead to any further change in $B_\textrm{p}$ (see the upper curve in Fig.~\ref{Valley_dep}b) despite the fact that, as already mentioned, the bare disorder potential continues to slowly increase with $\delta n$. This has also been confirmed in other samples. This set of experiments, therefore, allows us to conclude that valley polarisation itself is responsible for the reduction of $B_\textrm{p}$ observed in Fig.~\ref{Valley_dep}.

\begin{figure*}[t]
\includegraphics[width=2\columnwidth]{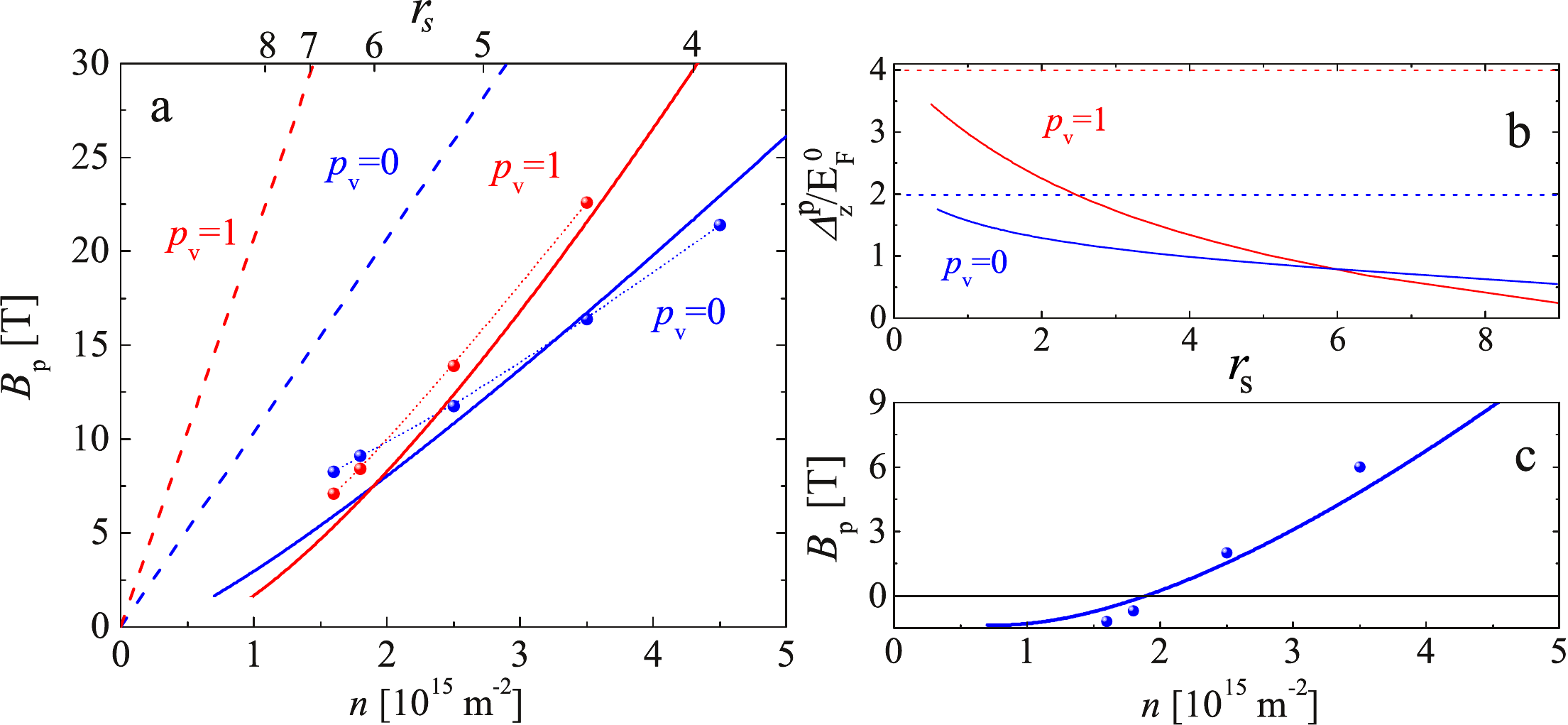}

\caption{\textbf{Comparison between experiments and quantum Monte Carlo simulations.} a) Density dependence of $B_\textrm{p}$. ($\cdot\cdot\bullet\cdot\cdot$) Experimental values, (\textbf{- - -}) non-interacting theory, (\textbf{---}) result of the quantum Monte Carlo simulation. b) Ratio of the Zeeman energy at full spin polarisation and $E_\textrm{F}^\textrm{0}$ as function of $r_\textrm{s}$. Dashed lines correspond to the non-interacting case while solid lines correspond to the prediction in the presence of interaction and disorder (The calculation are performed for the mobility of our sample). c) Density dependence of $\Delta B_\textrm{p}=B_\textrm{p}(p_\textrm{v}=1)-B_\textrm{p}(p_\textrm{v}=0)$. Experimental measurements ($\bullet$) and theory (\textbf{---}). \label{Prediction}}
\end{figure*}

\subsection*{Quantum Monte Carlo simulations}
The reduction in $B_\textrm{p}$ compared to single particle expectations is interpreted as resulting from electron-electron Coulomb interactions which favour spin alignment and this has already been confirmed theoretically
\cite{DePalo2005,DasSarma2005,DasSarma2006,Subasi2008,DePalo2009,DasSarma2009}. However, it is particularly challenging to make quantitatively reliable predictions for $B_\textrm{p}$. Even in the absence of disorder, Hartree Fock or random phase approximation (RPA) calculations do not capture the crucial role of correlations at low density and one has to resort to quantum Monte Carlo (QMC) simulations.\cite{DePalo2009} In our low-density intrinsically disordered system, the interplay of disorder and interactions further complicates the problem.

 Here, we use a Green's function quantum Monte Carlo approach, known to fully account for the effect of interactions in the presence of disorder\cite{Fleury2010}, to predict quantitatively the values of $B_\textrm{p}$ in the interacting 2DEG with and without valley polarisation (See Methods and ref.~\onlinecite{Fleury2010}). 

The energy per electron for a given spin polarisation state $p_\textrm{s}$ is given by:\cite{Fleury2010}
\begin{align}
E &=E_0+E_\textrm{p}p_\textrm{s}^2.
\end{align}
We calculate the energy $E_0$ of the spin unpolarised and the energy $E(p_\textrm{s}=1)$ of the spin polarised system to obtain the energy per electron needed to fully spin polarise the system $E_\textrm{p}=E(p_\textrm{s}=1)-E_0$.  
The predicted field of full spin polarisation is then calculated as $B_\textrm{p}=4E_\textrm{p}/(g\mu_\textrm{B})$.  The resulting $B_\textrm{p}$ as function of electron density can be easily compared to those extracted from the experiments with no adjustable parameter; the only input for the simulation being the amplitude $W$ of the bare disorder (which we can estimate from the peak mobility of the sample) and the electron density (See methods and Ref.~\onlinecite{Fleury2010} for more details on how disorder is taken into account in the model). This approach has been firmly validated by the successful quantitative comparison to the experimental measurements of seven different studies in Si at $p_\textrm{v}=0$.\cite{Fleury2010}. Here, we have extended those calculations to the case of $p_\textrm{v}=1$ using $\mu_{peak}=8000$ cm$^2\textrm{V}^{-1}\textrm{s}^{-1}$ measured in our sample as a single input parameter for both $B_\textrm{p}(p_\textrm{v}=0)$ and  $B_\textrm{p}(p_\textrm{v}=1)$ (see Methods for the determination of $\mu_{peak}$). Assuming that a single input parameter $\mu_{peak}$ determined at $p_\textrm{v}=0$ is enough to describe all the data seems reasonable since we have demonstrated that the bare disorder only varies weakly with $\lvert\delta n\lvert$. 

The result of the calculation is displayed in Figure~\ref{Prediction}a. In both valley-degenerate and valley-polarised cases the calculation predicts a much lower field of full spin polarisation than the single particle model. Importantly, the curves corresponding to $B_\textrm{p}(p_\textrm{v}=0)$ and $B_\textrm{p}(p_\textrm{v}=1)$ cross so that at low density the full spin polarisation is predicted to occur at a lower field in a valley polarised system. The full dependence of this effect from low to high interaction is illustrated in Fig.~\ref{Prediction}b. This figure shows the predicted ratio of the Zeeman splitting at full spin polarisation and the non-interacting Fermi energy $E_\textrm{F}^\textrm{0}$ as function of the interaction parameter $r_\textrm{s}=1/(\pi n)^{1/2}a_\textrm{B}$ (here $a_\textrm{B}$ is the Bohr radius) using the mobility of our sample. At low $r_\textrm{s}$ the behaviour is that of a non-interacting system. The effect of interaction becomes more and more important as $r_\textrm{s}$ increases and the curves corresponding to $p_\textrm{v}=1$ and $p_\textrm{v}=0$ eventually cross over at around $r_\textrm{s}=6$.

For a more quantitative comparison between experiments and theory, the experimental values of $B_\textrm{p}$ are plotted in Fig.~\ref{Prediction}a and in Fig.~\ref{Prediction}b the experimental difference $\Delta B_\textrm{p}=B_\textrm{p}(p_\textrm{v}=1)-B_\textrm{p}(p_\textrm{v}=0)$ is plotted together with the prediction. The theory describes the experiments very well with no adjustable parameters, demonstrating that the theory captures the essential physics behind the behaviour of $B_\textrm{p}$.

 \section*{Discussion}

\begin{figure}
 \includegraphics[width=\columnwidth]{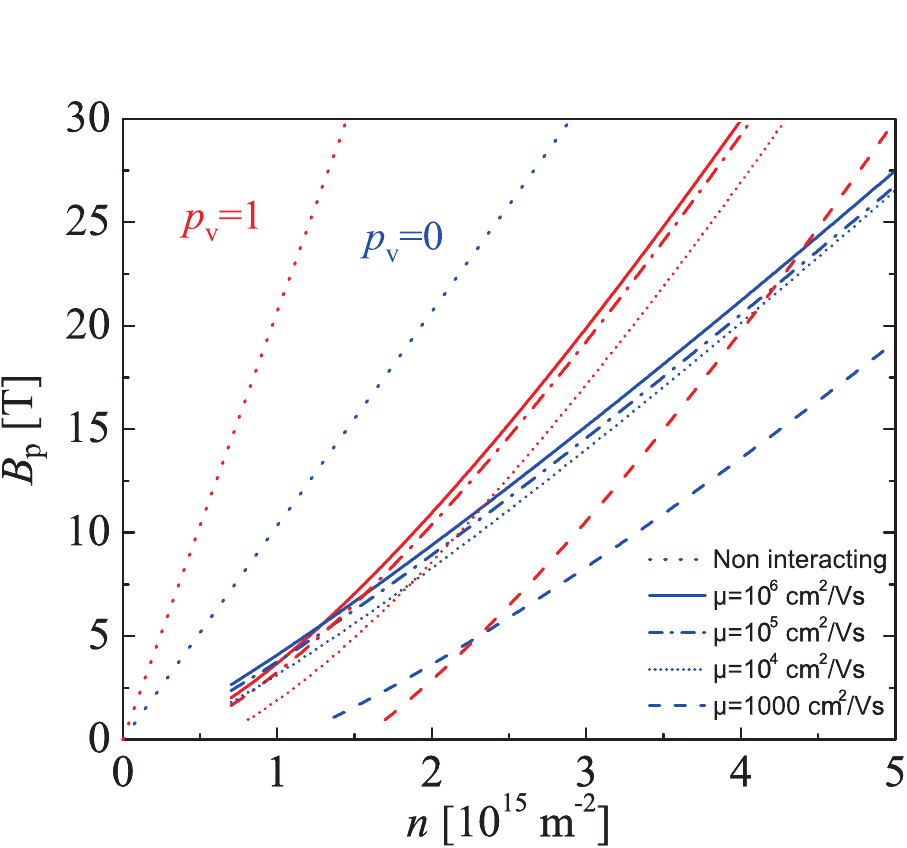}

\caption{\textbf{Density dependence of $B_\textrm{p}$.} Theoretical dependence of $B_\textrm{p}$ in a single particle 
picture and in presence of interactions for various values of disorder (quantum Monte-Carlo simulation). \label{Fig6}}
\end{figure}

As a first remark we should point out that the possibility of observing the new behaviour reported here results from the cooperative effect of disorder and interaction. Indeed, the calculations in the disorder free system indicate that electron-electron interactions are the leading effect in reducing the polarisation energies. Evidence for the crossing of $B_\textrm{p}(p_\textrm{v}=0)$ and $B_\textrm{p}(p_\textrm{v}=1)$ curves is also seen for clean systems (See Fig.~\ref{Fig6} and Fig.~2 in Ref.~\onlinecite{DePalo2009} where it is observed that the spin susceptibility enhancement $\chi/\chi_0(p_\textrm{v}=1)>2\chi/\chi_0(p_\textrm{v}=0)$ for large enough $r_\textrm{s}$, a feature reminiscent of the crossing of $B_\textrm{p}$ curves). However, in Si, this crossing is expected to occur at around $n=10^{15}$ m$^{-2}$, too low to be accessed experimentally in our valley tunable samples. By further enhancing the effects of electron-electron interactions disorder shifts by a small amount, all curves to larger densities, just enough to allow the observation of the new phenomenology. The weak dependence on disorder implies that our comparison is robust against errors in the determination of the bare disorder potential. This accounts for why $B_\textrm{p}$ is found to be independent of $\delta n<0$ in Fig.~\ref{negative_delta}d and justifies the use of a single input parameter to describe both $p_\textrm{v}=0$ and $p_\textrm{v}=1$. Nevertheless, including disorder is necessary to achieve quantitative comparison. This effect of disorder may explain why, in previous experiments in AlAs at similar values of $r_\textrm{s}$, only the enhancement of spin susceptibility and not the reduction of $B_\textrm{p}$ with valley polarisation was observed\cite{Shayegan2004}.  Indeed, in AlAs the mobility was five times larger than in our samples and therefore the crossing would be shifted to larger $r_\textrm{s}$. However, the crossing must have been approached very close since the criteria $\chi/\chi_0(p_\textrm{v}=1)>2\chi/\chi_0(p_\textrm{v}=0)$ is almost reached in Fig.~3 of Ref.~\onlinecite{Shayegan2004}. In addition, we should also point out that the comparison between QMC calculations and experimental results in valley tunable AlAs is complicated by the mass anisotropy in the system as concluded in Ref.\onlinecite{Shayegan2007b}. The situation is more simple in narrow AlAs quantum wells where effective mass is isotropic\cite{vakili2004} and QMC works well.\cite{DePalo2005}

As a second remark we note that our study further confirms that QMC is an appropriate theory to predict the polarisation energies of two-dimensional electron gasses. Previous studies have shown that the disorder free QMC describes correctly the measurements in AlAs (apart from $p_\textrm{v}=1$ in valley tunable AlAs with anisotropic effective mass) and GaAs if the finite thickness of the system is included\cite{DePalo2005,DePalo2009}. We have also shown in Ref.~\onlinecite{Fleury2010} that the QMC including disorder describes the available experimental data in silicon at $p_\textrm{v}=0$. Therefore, QMC is able to determine the polarisation energies of most investigated samples without adjustable parameters. 

As a concluding remark we now discuss our result in a more general context. Early QMC simulations\cite{Conti1996} on clean 2DEGs at $B=0$ T showed that the energy of the spin and valley polarised system becomes lower than that of the two component system in the region $r_\textrm{s}\gtrsim20$ before Wigner crystallisation at $r_\textrm{s}\sim34$. A ferromagnetic instability had therefore been expected in the valley polarised system. In contrast, and in qualitative contradiction with results from Hartree-Fock calculations, no such instability was observed in the QMC simulations for the valley and spin degenerate system which was found to be the stable phase all the way up to Wigner crystallisation at $r_\textrm{s}\sim42$ (See Fig.~1 in Ref.~\onlinecite{Conti1996}). Our experiments and quantum Monte Carlo simulations confirm and demonstrate that this scenario remains valid in real disordered systems, rather than it being only a special case of hypothetical disorder-free systems. Results from simulations displayed in Fig.~\ref{Prediction}a show that $B_\textrm{p}(p_\textrm{v}=1)$ is approaching 0 at a finite electron density indicating a ferromagnetic instability in the valley polarized system. The instability occurs at a lower $r_\textrm{s}$ in our disordered system compared to the clean one\cite{Conti1996}. In contrast, no sign of such instability is seen in the curve for $B_\textrm{p}(p_\textrm{v}=0)$ down to the lowest electron densities we have studied. Experimentally, the mobility of our sample obviously does not allow us to reach the very low densities necessary for the observation of spontaneous spin polarisation or Wigner crystallisation. Yet, the crossing of the curves $B_\textrm{p}(p_\textrm{v}=0)$ and $B_\textrm{p}(p_\textrm{v}=1)$ provides strong experimental evidence of the higher stability of the spin-valley degenerate system because they demonstrate that interaction-induced spin alignment is much less efficient in the valley unpolarised system than in the valley polarised system. The strong enhancement of spin susceptibility with valley polarisation in AlAs\cite{Shayegan2004} also supports this interpretation. The excellent agreement between the theory and our experiments suggests that the result can be extrapolated to cleaner spin-valley degenerate systems with greater interaction. In those systems, in the absence of ferromagnetic instability, we anticipate the presence of a strongly correlated electron liquid. This may result in a rich physics which might soon be accessed exploiting recent developments in high mobility silicon systems\cite{Tsui2012,Kane2014}.

\section*{Methods}
\subsection*{Samples and control of valley polarisation}
The samples consists of a SiO$_2$/Si(100)/SiO$_2$ quantum well of nominally 10nm thick silicon with front- and back-gate oxide thicknesses of 75 nm and 380 nm respectively. The fabrication proceedure of
this sample is described in Ref.~\onlinecite{Takashina2004}. A degenerately phosphorus-doped polysilicon layer was used as front gate while the substrate was used as a back gate.

It is well known that valley degeneracy can be lifted at the (001) Si/SiO$_2$ interface in Si MOSFETs and that the valley splitting can be increased by increasing the out-of-plane electric field, which in traditional MOSFETs can be controlled by changing the substrate bias \cite{AndoFowlerStern1982}. The magnitude of this valley splitting is found also to depend on the way in which the Si-SiO$_2$ interface is prepared, and the use of a buried-oxide interface using SIMOX (Separation by IMplantation of Oxygen) technology \cite{Takashina2004} allows us to enhance the valley splitting up to tens of meV \cite{Takashina2006}. The coupling responsible for the bare single-particle splitting is induced predominantly by the large interface electric field\cite{Saraiva2009,Saraiva2011} but there remain uncertainties as to the exact microscopic details that give rise to the particularly large values in SIMOX buried-oxide interfaces\cite{Saraiva2009,Saraiva2011,Saraiva2010}.

Experimentally, the valleys splitting is determined by fitting Shubnikov de Haas oscillations with an empirical expression for valley splitting\cite{Takashina2006}:
\begin{equation}
\label{Deltav}
 \Delta_\textrm{v}=\alpha \delta n
\end{equation}
where, 
$\delta n$ is an empirical measure of the out-of-plane electrostatic potential asymmetry controlled by front and back gates: 
\begin{equation}
\delta n = n_{\mathrm{B}}-n_{\mathrm{F}}
\end{equation}
where $n_{\mathrm{F}}$ and $n_{\mathrm{B}}$ are electron densities contributed by respective gates. Both $n_{\mathrm{F}}$ and $n_{\mathrm{B}}$ can take positive or negative values where negative values represent a density reduction due to a depleting bias from the corresponding gate so that the total electron density is given by
\begin{equation}
 n = n_{\mathrm{B}}+n_{\mathrm{F}}.
\end{equation}
The numerical factor $\alpha$ which is of the order of 0.5 meV/$10^{15}$m$^{-2}$ determines how much the valley splitting changes with $\delta n$ when $\delta n$ is positive. That is, when the quantum well is biased in such a way that the electrons are pulled towards the back (SIMOX buried-oxide) interface. On the other hand, when $\delta n$ is negative, the electrons are pushed against the front interface, which is formed by standard thermal oxidation, where we find the valley splitting to be negligibly small. Thus, by pressing the electrons against the buried-oxide interface (positive $\delta n$) we can increase the valley splitting continuously, and independently control the electron density $n$. The out-of-plane potential necessarily affects the disorder, however, but the effects of this can be independently examined by applying a negative $\delta n$ for which there is no valley splitting\cite{Takashina2013}.

For fitting the Shubnikov de Haas oscillations, we fix the perpendicular magnetic field and compare the valley splitting $\Delta_\textrm{v}$ against the cyclotron energy $\hbar \omega_c$, or more accurately, we map the number of occupied Landau levels of the two valleys as function of $(\delta n, n)$. Straightforwardly applying the single-particle model only yields a value for $\alpha m_\textrm{b}$ but not $\alpha$, in the same manner as coincidence experiments under tilted field only provide values for $g m_\textrm{b}$ and not $g$. The bare effective mass only provides a crude conversion of the valley splitting to an energy scale\cite{Takashina2006}, and would represent a good measure of the valley splitting in the absence of interactions that alter the effective density of states. 

Valley polarisation is then estimated at full spin polarisation from the equation:
\begin{eqnarray}
	p_\textrm{v} &=& \frac{\Delta_\textrm{v} D_\textrm{0}}{n}; \nonumber\\
	p_\textrm{s} &=& 1, \nonumber\\
\label{Valleypol}
\end{eqnarray}
The determination of valley polarisation does not, therefore, rely on separating $\alpha$ and $m_\textrm{b}$ since in Eqs.~\ref{Valleypol}, $\Delta_\textrm{v}$ never appears on its own but always as a product with $D_\mathrm{0}$. It follows that this equation remains a valid method of determining the valley polarisation even in the presence of strong interactions. See Supplementary Discussion for more details.

\subsection*{Electrical Measurements}
The samples were cooled in a Variable Temperature Insert with a base temperature of 1.6 K inserted into a 30 T resistive magnet. A standard four-terminal lock-in technique was used to measure the resistivity $\rho_{xx}$ and $\rho_{xy}$ of a sample with a Hall-bar geometry. The current was kept below 5 nA to avoid electron heating. The samples were aligned parallel to the applied magnetic field with an \textit{in-situ} rotator, eliminating the Hall resistance.

\subsection*{Determination of the mobility}

The comparison of the theory to the experimental data requires the determination of the mobility of the sample in the Drude regime (conductivity  $>>e^2/h$, see the next section). In practice one should measure the resistance at high density to obtain this quantity. However, this is not straightforward in our case because for densities larger than $n=5\times10^{15}$ m$^{-2}$, electrons start to experience scattering from localized states in the upper spatial subband which become populated\cite{Niida2013}. Fortunately, the mobility at high density in the absence of the influence of the upper spatial subband can still be determined in our sample. To do so, we exploit the fact that pressing the electron gas to the front interface at negative $\delta n$ does not increase valley splitting but narrows the effective width of the out-of-plane electronic wavefunction. This pushes the upper spatial subband to higher energies and suppresses its influence. This is illustrated by the solid white line in Fig.~\ref{Fig4} which demarcates the onset of occupation of the upper subband. As the out-of-plane electric bias ($\delta n$) is increased, the density $n$ at which it starts to fill increases, reflecting the increasing confinement energy\cite{Takashina2006}. The suppression of scattering by localized states in the upper spatial subband is evidenced by the reduction of resistance with $\delta n$ for densities about $10^{16}$ m$^{-2}$. For densities relevant to the present study, scattering by localized states in the upper subband is absent as shown by the blue constant resistance contour in Fig.~\ref{Fig4} which is parallel to the constant density line for all negative $-2\times 10^{16} \text{m}^{-2}<\delta n<0$. For even larger electric fields $\delta n <-2\times 10^{16}\text{m}^{-2}$ the electron gas experiences the roughness of the front interface\cite{Niida2013} and the resistance starts to increase so that the iso-resistance lines deviate from constant density lines. We conclude that the mobility at high density and in absence of the influence of the upper spatial subband can be estimated in the regime $-2\times 10^{16} \text{m}^{-2}<\delta n<-1\times 10^{16} \text{m}^{-2}$. Therefore, we estimated $\mu$ for $n=10^{16}$ m$^{-2}$ and $\delta n=-1.5\times 10^{16}\text{m}^{-2}$ marked by a star in the Fig.~\ref{Fig4}, where $\rho=760 \Omega$ leading to $\mu=8000$ cm$^2\textrm{V}^{-1}\textrm{s}^{-1}$.

\subsection*{Quantum Monte Carlo simulations}
 
The model we consider is a generalization of the Anderson model to the many-body problem (see Ref.~\onlinecite{Fleury2010,Waintal2006} for details). The system is made of $N$ spin up/down electrons with Coulomb interaction on a disordered lattice of $L_x\times L_y$ sites. The electrons either populate a single valley ($p_\textrm{v}=1$) or are equally split up into two degenerate valleys ($p_\textrm{v}=0$). Their spin configuration sets the value of the spin polarisation $p_\textrm{s}$. Formally, the spin and valley degrees of freedom are treated strictly in the same way as an internal electronic degree of freedom. In the continuum limit $\nu\equiv N/L_xL_y\ll 1$ (where lattice effects are negligible) and in the thermodynamics limit $N\gg 1$, the physics of a given $(p_\textrm{s},p_\textrm{v})$ configuration is entirely controlled by two dimensionless parameters $r_\textrm{s}=m_\textrm{b} e^2/(4\pi\epsilon\hbar^2\sqrt{\pi n})$ ($m_\textrm{b}=0.19m_e$ effective mass, $e$ electron charge, $\epsilon=7.7\epsilon_0$ dielectric constant) and $1/k_Fl$ ($k_F$ Fermi momentum, $l$ mean free path, both taken for the spin- and valley-degenerate system) which characterize respectively the interaction strength and the disorder strength in the system. Experimentally, it is difficult to estimate $k_F l$ for low density systems because conductivity is no longer a good estimate of disorder. We overcome this issued by observing that for white noise disorder of amplitude $W$ (as assumed in our model), $k_F l\propto n$. 
Therefore, $\eta \equiv r_\textrm{s}\sqrt{k_Fl}$ does not depend on $n$ and depends only on $W$\cite{Fleury2010}. One can estimate $\eta$ in the high density regime where electronic interactions are negligible. In that regime, the conductance $g$ of the system is $g=(2e^2/h)k_Fl$ which gives $\eta=\sqrt{\mu}e^{3/2}m/(4\pi\epsilon\hbar^{3/2})$ where $\mu=g/(en)$ is the mobility of the sample. Thus, the two input parameters $r_\textrm{s}$ and $k_Fl$ of our model can be estimated from the mobility of the sample measured at high density and from the experimental values of electronic densities. 

 We use the Green's Function Monte Carlo method~\cite{Trivedi1990} in the fixed-node approximation to compute the energy per particle $E(p_\textrm{s})$ of the ground state of our model at zero temperature~\cite{Fleury2010}. The polarisation energy $E_p$ is deduced from $E_p=E(p_\textrm{s}=1)-E(p_\textrm{s}=0)$ and averaged over 50 to 200 samples depending on the disorder strength. To extrapolate data at the thermodynamic and continuum limit, finite $N$- and $\nu$-effects are carefully investigated. We thus observed large but controlled lattice effects without interaction that disappear as interaction is switched on ($r_\textrm{s}\gtrsim 0.5$). Small finite size effects in $N$ are also present but they rapidly fade with the disorder amplitude. The resulting extrapolated data, as well as their fits given below, are finally obtained with a precision of the order of $\pm 0.02 E_\textrm{F}^0$ (for $k_F^0l_0\geq 1.5$) and $\pm 0.04 E_\textrm{F}^0$ (for $0.3<k_F^0l_0<1.5$) at $p_\textrm{v}=0$, and roughly twice larger at $p_\textrm{v}=1$.\\

 Without interaction ($r_\textrm{s}=0$), our data are in perfect agreement with the second-order perturbative formula $E_p/E_\textrm{F}^0=1/2\,[1]+\log2/(\pi k_F^0l_0)$ for $p_\textrm{v}=0$ [$p_\textrm{v}=1$], at least for weak disorder ($k_F^0l_0 \gtrsim 0.4$). In the presence of (even small) interaction, first the effect of disorder is reversed making easier the spin polarisation of the system and second, the $1/k_F^0l_0$-correction to $E_p$ is no longer valid (except for tiny disorder). We find that our $E_p$ data are very well described by the following formula,
\begin{equation}
\label{eq_Ep}
E_p(k_F^0l_0,r_\textrm{s})=E_p^{cl}(r_\textrm{s})+\frac{\beta(r_\textrm{s})}{\sqrt{k_F^0l_0}}\,E_\textrm{F}^0+A_5\,E_\textrm{F}^0\,,
\end{equation}
where the polarisation energy of the clean system $E_p^{cl}$ and the parameter $\beta$ are both fitted with Pad\'{e} approximates,
\begin{align}
E_p^{cl}(r_\textrm{s})&=\frac{A_0+A_1\,r_\textrm{s}}{A_2+A_3\,r_\textrm{s}+A_4\,r_\textrm{s}^2}\,E_\textrm{F}^0   \label{eq_Epcl} \\    
\beta(r_\textrm{s})&=\frac{B_0+B_1\,r_\textrm{s}^2}{B_2+B_3\,r_\textrm{s}+B_4\,r_\textrm{s}^2}\,,    \label{eq_alpha}
\end{align}
the fitting parameters $A_i$ and $B_i$ being given in Table~\ref{tab_fitPade}. We note that Eq.~\ref{eq_Epcl} for $E_p^{cl}$ is in very good agreement with 
previous Quantum Monte Carlo calculations performed for the valley-polarized system~\cite{Attaccalite2002} and the valley-degenerate system~\cite{DePalo2009}. 
Equation~\ref{eq_alpha} -- and in particular the fact that $\beta$'s sign flips at $r_\textrm{s}\approx 0.3$ (for $p_\textrm{v}=0$) and $r_\textrm{s}\approx 1.2$ (for $p_\textrm{v}=1$) -- mainly depicts 
the opposite effects of disorder at very weak and stronger interactions. The last parameter $A_5$ adjusts the origin of the linear disorder correction in 
$1/\sqrt{k_F^0l_0}$, to roughly take into account the actual quadratic correction in $1/\sqrt{k_F^0l_0}$ at weak disorder. At extremely weak disorder, $1/(k_F^0l^0)\lesssim 0.04$, $A_5$ shall be taken equal to 0 and $E_p$ evaluated by $E_p^{cl}$. We point out that Eqs.~\ref{eq_Ep}-\ref{eq_alpha} are no more than one simple way to report our data, valid (at least) for $0.25\,[0.5]\leq r_\textrm{s}\leq 10$ (at $p_\textrm{v}=0$ [$p_\textrm{v}=1$]) 
and $k_F^0l_0\geq 0.3$ as long as the output $E_p$ is positive.\\
\indent Deducing the polarisation magnetic field $B_\textrm{p}$ from $E_p$ is straightforward, once noticing that the energy $E$ of the ground state is quadratic in $p_\textrm{s}$,
\begin{equation}
\label{eq_E_vs_ps}
E(p_\textrm{s})=E(0)+E_p p_\textrm{s}^2\,.
\end{equation}
This statement is obvious in the absence of disorder and interaction where we have $E_p=E_\textrm{F}^0/2\,[E_\textrm{F}^0]$ for $p_\textrm{v}=0$ [$p_\textrm{v}=1$]. Numerically, it turns out to 
remain valid with good precision for intermediate disorder and interaction strength ($0\leq r_\textrm{s} \leq 10$, $k_F^0l_0>1$). In particular, Eq.~\ref{eq_E_vs_ps} 
is satisfied in the disorder and interaction regime explored in the present experiment. Then, when an in-plane magnetic field $B$ is applied, a Zeeman term 
$-g\mu_\textrm{B} B p_\textrm{s}/2$ has to be added to the right hand side of Eq.~\ref{eq_E_vs_ps}. Minimizing the energy $E$ with respect to $p_\textrm{s}$ gives the spin polarisation 
of the system at zero temperature, $g\mu_\textrm{B} B/(4E_p)$, from which we get $B_\textrm{p}=4E_p/(g\mu_\textrm{B})$.\\
Figure~\ref{Fig6} presents the result of the numerical calculations of the magnetic field of full spin polarisation for various values of disorder. This figure shows that even in the presence of weak disorder ($\mu=10^6$ cm$^2\textrm{V}^{-1}\textrm{s}^{-1}$), the magnetic field of full spin polarisation is expected to be much lower than in the non-interacting picture.

\begin{table}[h]
\centering
\setlength{\extrarowheight}{2pt}
\begin{tabular}{|c|c|c|c|c|c|c|}
\hline
$i$ & 0 & 1 & 2 & 3 & 4 & 5 \\
\hline
$A_i(p_\textrm{v}=0)$ & 27.93 & 9.83 & 56.5 & 46 & 1.77 & 0.019 \\
$A_i(p_\textrm{v}=1)$ & 23.4 & -0.5 & 24.4 & 7.75 & 0.27 & 0 \\
$B_i(p_\textrm{v}=0)$ & 2.70 & -24.81 & 42.4 & -11.5 & 246.9 &  \\
$B_i(p_\textrm{v}=1)$ & 54.28 & -39.30 & 348 & 170.5 & 267 &  \\
\hline
\end{tabular}
\caption{\textbf{Coefficients for the estimation of polarisation energies.} $A_i$ and $B_i$ parameters of Eqs.~\ref{eq_Epcl} and~\ref{eq_alpha} for the valley-degenerate ($p_\textrm{v}=0$) and valley-polarized ($p_\textrm{v}=1$) system.}
\label{tab_fitPade}
\end{table}

The code to perform this simulations has been parallelised and ported on CEA Computing Center for Research and Technology (CCRT) massive parallel clusters. About 100000 CPU hours have been required for the present study.
The code is available upon request.

\subsection*{TEM images}
The High Angle Annular Dark Field scanning TEM images were measured using a probe-aberration corrected FEI Titan microscope operated at 200 kV. A 100-nm-thick specimen was prepared by focused ion beam milling at 5 kV to reduce the surface damage. The HAADF STEM images are sensitive to Z-contrast and the vertical bright “dumb-bell” structures are typical of aberration corrected images of silicon samples oriented in the $< 1 1 0 >$ direction and show silicon atoms separated by 1.36 Angstroms in projection.

\section*{Acknowledgements}
We acknowledge the support from the 7th framework programme ”Transnational Access” (contract N 228043-EuroMagNET II 
Integrated Activities Ref. GSC09-210) and the support from the CEA-CCRT supercomputing facilities. 
VTR acknowledges support from the Nanoscience Fundation of Grenoble. XW is founded by the ERC consolidator grant MesoQMC. DC is supported by the ERC starting Grant Holoview. KT and DT are supported by the EPSRC of the UK(EP/I017860/1). AF acknowledges the support by the Funding Program for Next Generation World-Leading Researchers of JSPS (GR103). YH acknowledges support from JST programs. 

\section*{Authors contribution}
VTR, BAP and KT designed the experiments. AF fabricated the samples. VTR, BAP, YN, DT and KT performed the transport measurements which were discussed by all the authors. 
XW and GF performed the quantum Monte Carlo simulations. DC performed TEM measurements. VTR, KT, GF and XW co-wrote the manuscript and all authors commented on it. KT coordinated the collaboration.

\section*{Competing financial interests}

The authors declare no competing financial interests.


\end{document}